\documentstyle[12pt]{article}
\begin{document}
\thispagestyle{empty}
%\preprint{Usach 950301, hep-th 9503xxx}
\def\D{{\cal D}}
\def\12{{1\over 2}}
\def\st{\displaystyle} 
\def\bb{\begin{equation}} 
\def\ee{\end{equation}}
\def\G{G[x_2, x_1]} 
\def\bn{\begin{eqnarray}} 
\def\en{\end{eqnarray}} 
\title{Quantum Mechanics on Multiply Connected Manifolds \\ 
with Applications to Anyons in One and Two Dimensions}
\author{{\it J. Gamboa} \\ 
{\it Departamento de F\'{\i}sica, Universidad de Santiago de Chile}\\ 
{\it Casilla 307, Santiago 2, Chile}}
\date{}
\maketitle
 
\pagestyle{empty}

\section{Introduction} 

Physics defined on multiply connected manifolds is an old topic 
in theoretical physics. In the context of the path integral
formalism it was studied by the first time by Schulman \cite{schulman} in
1968 and rigourely formulated by Laidlaw and de Witt
\cite{dewitt} and Dowker \cite{dowker} in 1971.

The central point is that in a multiply connected manifold the
paths have differents weights in the sum over histories and the
problem -it does not exist in the \lq \lq standard" quantum
mechanics- is how to define a quantum theory taking in account
this fact.

The problem of to how define a quantum theory on a topologically 
non-trivial manifold is not an academic problem but since it
finds experimentally realizable systems such as the
Aharanonov-Bohm effect and the anyons that could an explanation to the
quantum Hall effect and, maybe, to high temperatures
superconductivity. 

The purpose of these lectures will be to explain some aspects of
the quantum theory defined on multiply connected manifolds in
the context of the path integral formulation and the
applications that these ideas find in anyon physics in one and
two dimensions. 

Let us start in section 2 explaning some examples that involve 
non-trivial topological aspects; this section does not
involve calculations and the unique purpose is to introduce
several useful concepts. In section 3, we introduce the formal
definition of a multiply connected manifold. In Section 4, path 
integrals on arbitrary manifolds. In section 5, several
applications are studied and in section 6, the conclusions are
given. 
     
\section{Systems Defined on Non-Trivial Topological
Manifolds} 

Let us start discussing the most popular example of a quantum
theory defined on a multiply connected manifold, namely the 
Aharonov-Bohm effect (AB).

The AB effect consists in the following experimental arrangement
\begin{figure}
\vspace{4cm}
\caption{The solenoid has infinite lenght and inside a
constant magnetic field, it assumes also that the solenoid is
impenetrable.} 
\end{figure} 

%The electrons can follow infinite paths   
\begin{figure}
\vspace{5cm}
\caption{Some of the the infinites path of the electron} 
\end{figure}

The important question is that theoretically one expect that in
the screen interference lines can be observed such as in a
diffraction experiment and the lines be dependent only on the
magnetic field  that is inside of the solenoid. 

This example tell us that the electromagnetic potentials -that
classically are unobservable- are quantum mechanically
responsible for the observability of the interference pattern.
The experimental question relative to the AB effect was only solved
with a serie of experiments performed by Tonomura and
collaborators in the beginning of the eighties\cite{tonomura}
... twenty five years after of the Aharonov and Bohm prediction.

From a theoretical point of view, one can see the AB effect as
phenomenon that occurrs because the $\Re^2$ manifold (that
is the plane where the paths live) has a point removed (the
point where the solenoid is) and, as a consequence, the
configuration space of this system is $\Re^2 -\{0\}$. 

The AB effect is an example of a mechanism that appears in many
examples of recent physics, one of them is the problem of 
two-anyons. 

In order to explain this problem, let us consider the motion of
two non-relativistics particles moving on a plane. The motion is
regular everywhere except in the point where the particles
collide. 

The colision condition in the point $x_1=x_2$ is equivalent to
the replacement 
\begin{equation} 
 \Re^2 \rightarrow  \Re^2 -\{0\}, \label{manifold} 
\end{equation} 
and, in consequence, the manifold (configuration space) has also
a point removed. 

One can see formally this example as a similar phenomenon to the
AB effect, each particle has atached a flux-tube and in the case
of the two particles, one can exactly map it to the AB effect.

Of course, there are questions that one should give an answer; 
what is the analogue of the magnetic field for the case of two 
particles?, how to implement technically this fact?, etc.. 

There is also another problem closely related with the previous
ones, namely cosmic strings.  The cosmic strings are solutions
of the Einstein field equations when matter like point is
present, the solution is 
\begin{equation} 
dS^2 = dt^2 - r^2 {d\phi^{'}}^2 - dr^2 - dz^2, 
\end{equation} 
where $\phi^\prime$ is the defect angle.

The cosmic strings are generally assumed to be singularities
that remained after the formation of our universe and could be
experimentally detectable. 

The manifold when one project to a plane is again 
$\Re^2 -\{0\}$ . 

In the next sections we will try to formalize these examples
developing appropriate techniques of calculation. 

\section{Rudiments of Homotopy Theory}

The configuration space where one compute the propagator of a free
particle is an example of a simply connected manifold. When we
give two points of the manifold one can draw infinite paths
between these points which are topologically equivalent.

The word \lq deformable\rq has a technical connotation for a
mathematical operation called homotopy transformation that we
will define below\cite{nasch}. 

The idea of a homotopy is the following; we will say that two
curves are homotopically equivalent if it is possible to deform
continuosly one into the other, in other words, two continuous applications 
$f$ and $g$ of the space $X$ to the space $Y$, $g: X\rightarrow Y$
are homotopic (simbolically $ f\sim g$) if it exists a
continuous function $F: X\times I \rightarrow X$, where $I$ is the closed
interval $[0,1]$, such that 
\begin{equation} 
F(x,t)\vert_{t=0} = f(x), \nonumber 
\end{equation} 
and 
\begin{equation} 
F(x,t)\vert_{t=1} = g(x), \nonumber 
\end{equation} 
with $(x,t)\in X$. 

Is clear that the idea of homotopy define a class of equivalence
between applications, {\it i.e.}, 

$i)$ $f\sim f$, 

$ii)$ $f\sim g$ $\Rightarrow$ $g\sim f$, 

$iii$ $f\sim g$ and $g\sim h$ $\Rightarrow$ $f\sim h$, 

$\forall$ continuous function $f,g$ and $h$. 

If $G$ is the space of the all continuous applications between
$X$ and $Y$, then a relation of equivalence has the property of 
decomposing the space in classes of equivalence or disjoint sets of
functions homotopically equivalent. 

If the functions $g$ and $f$ are homotopic, then they belong to
the same class of homotopy, otherwise they are non-homotopically
equivalent. We will denote the homotopy class by $[\alpha]$
where the set $[\alpha]$ is the set of paths that are
homotopically equivalent. In the case of the AB effect in fig.
2 the path $1$ and $2$ belong to the same class of homotopy. 

Now, we will restrict only to the applications that are closed
curves or loops; we will say that the loops $\alpha$ and $\beta$
with basis in $x_0$ ({\it i.e} the point where the extremes
coincide) are equivalent if there exists a function $H: I\times I
\rightarrow X$ such that $H(t, 0) = \alpha$, $H(t^{'}) = \beta$
and $H(0,s) = H(1,s)= x_0$ $\forall s \in I$. 

The function $H(s,t)$ is a homotopy, then if $\alpha , \beta$
and $\gamma$ are loops with basis in $x_0 \in X$  then 

$i)$ $\alpha \sim \alpha$, {\it i.e.} any loop is equivalent
itself. 

$ii)$ If $\alpha \sim \beta$  $\Rightarrow$ exist a homotopy 
$H: I\times I \rightarrow X$ with $H(t, 0) = \alpha$, $H(t,1)=
\beta$ and $H(0,s) = H(1,s)= x_0$. 

$iii)$ $\alpha \sim \gamma$ if $\alpha \sim \beta$ and $\beta
\sim \gamma$. 

It is possible to define a homotopy $L(t,s)$ between $\alpha$ and
$\gamma$ as follows 
\begin{equation} 
L (t,s) = \left\{ \begin{array}{ll} 
H(s,2t) & \mbox{ $0\leq t \leq {1\over 2}$} \\ 
H(s, 2t - 1) & \mbox{ ${1\over 2} \leq t \leq 1$} 
\end{array} \right.   
\end{equation}
and in consequence $\alpha \sim \gamma$. 

One can think of a more tangible example considering the AB
effect \lq \lq joining" the two extremes in fig. 2. Observing
fig. 2 one see that there are paths that can be classified by a 
\lq \lq topological invariant" $n$ (winding number). 

In general the loops can be summed and the resulting sum is
another loop that link $\sum n$ times the hole. The set of all
loops is a group that is isomorphic to the integer group, however in order
to implement this fact it is necessary to define the product of
loops. 
The definition is the following; let $\alpha$ and $\beta$ curves
in $X$ with $\alpha (1) = \beta (0)$, then the product $\ast$ 
of curves is  
\begin{equation} 
 (\alpha \ast \beta) (t) = \left\{ \begin{array}{ll} 
H(2s,t) & \mbox{$0\leq s \leq {1\over 2}$ } \\ 
H(2s -1,t) & \mbox{${1\over 2}\leq s \leq 1$} 
\end{array}\right.   
\end{equation} 
Once these definitions are given, one can show that the set of
all the homotopy classes of loops $\{ [\alpha ]\}$ with basis in
$x_0$ of $X$ is a group and is denoted by $\pi_1 (X, x_0)$ and
is formally equivalent to 
\begin{equation} 
\pi_1 (X, x_0) = \{ [\alpha ]\}. \nonumber 
\end{equation} 

The set $\pi_1$ endowed of the operation $\ast$ defines the
first homotopy group or fundamental group. The group $\pi_1$ is
the first of an infinite set of $(n>0)$ higher order homotopy
groups, $\pi_1$ eventually can be non-abelian while the higher
homotopy groups are all abelian. 

\section{ Path Integrals on Multiply Connected Manifolds} 

In this section we will introduce the concept of path integrals
on multiply connected manifolds. Let us start considering path 
integrals on simply connected manifolds and the most
simple application, namely, the free non-relativistic particle; 
this example is a warm up exercise and will be
useful when the path integral on a multiply connected manifold 
is considered at the end of this secton. 

The idea of the path integral consist to sum on all the paths
between the initial and final points $A$ and $B$. 
The propagation amplitude between these points is equivalent to
the computation of the formal sum
\begin{equation} 
G [B, A] \sim \sum_{paths} \lq \lq something".
\end{equation}

The previous expression has two difficulties, firstly has the
technical problem of how to define the sum between paths and second
it has the physical problem of how to define $\lq \lq something"$.
This second problem is equivalent to postulate the Schr\"odinger
equation in the conventional quantum mechanics and it is
equivalent to make the replacement 
\begin{equation} 
\lq \lq something"  \rightarrow e^{{i\over {\hbar}} S}, 
\end{equation} 
where $S$ is the action. 

The first problem is technically more difficult and, in essence,
their solution consists in replacing (for details see {e.g}
\cite{faddeev}) 
\begin{equation} 
\sum_{paths} e^{{i\over {\hbar}} S} \rightarrow \int \prod_t dx(t) 
e^{{i\over {\hbar}}\int dt L(x, {\dot x})}, 
\end{equation} 
where $L(x, {\dot x})$ is the lagrangian of the system. 

In general although one can give a discretization prescription, the
physical quantities are well defined only giving correctly the
boundary conditions. Thus, if one is interested computing
the propagator of a particle, we must give the boundary conditions
\begin{equation} 
x(t_1) = x_1, \,\,\,\,\,\,\,\,\,\,\,\,\,\,\,\,\,\,\,\,\,\,\, 
x(t_2) = x_2, \label{bc}
\end{equation} 
and then the expression 
\begin{equation} 
G[x_2, x_1] = \int {\cal D} x(t) e^{{i\over h} S}, \label{propa}
\end{equation} 
plus (\ref{bc}) defines the propagation amplitude or propagator of
the system, here ${\cal D} x(t) = \prod_{t} d x(t)$.

One can verify explicitly how work out these ideas considering
explicitly the most simple example, namely the motion of a free
non-relativistic particle in one dimension described by the lagrangian 
\begin{equation} 
L = {1\over 2} {\dot x}^2. 
\end{equation} 

We are interested in computing the propagation amplitude
$G[x_2,x_1]$ with the boundary conditions (\ref{bc}). In order to compute 
\bb 
\G = \int \D x(t) e^{{i\over h} \int_1^2 dt \12 {\dot x}^2},
\label{formula} 
\ee 
one start making the following change of variables 
\bn  
x(t) &=& x_1 + {\Delta x \over \Delta t} (t - t_1) + y(t)
\nonumber \\ 
&=& x_{cl} + y(t), \label{eq}
\en 
where $x_{cl}$ is the solution of the classical solutions of 
equation of motion ${\ddot x} = 0$ and $y(t)$ is a quantum fluctuation that,
by consistency, satisfy the boundary condition 
\bb 
y(t_1) = 0 = y(t_2). \label{homo} 
\ee 
When (\ref{eq}) is replaced in (\ref{formula}) one find that 
\bb 
\G = e^{i{{(\Delta x)}^2\over 2\Delta t}} \int \D y(t) 
e^{{i\over 2}\int_1^2 dt y(-\partial^2_t) y}. \label{gauss} 
\ee 

The integral in $y$ is gaussian and the result of the
integration is 
\bb 
\det (-\partial^2_t)^{-{\12}}, 
\ee 

Now one should compute the determinant, the procedure of 
calculation is the following: one start by solving the eingenvalue
equation 
\bb 
\partial^2_t  \psi_n = \lambda_n \psi_n, \label{eigen} 
\ee 
with Dirichlet boundary conditions and afterwards one use the
formula 
\bb 
\det (-\partial^2_t) = \prod_n \lambda_n. \label{determina}
\ee 

Using $\psi_n (t_1) = 0 = \psi_n (t_2)$, we find that $\lambda_n
= {(n\pi /\Delta t)}^2$ and (\ref{determina}) becomes 
\bb 
\det (-\partial^2_t) = \prod_{-\infty}^{+\infty} {(n\pi /\Delta
t)}^2, \label{diverge} \ee 
however, (\ref{diverge}) is divergent. 

In order to give sense to the divergent expression
(\ref{diverge}), one regularize appropiately this product;
firstly one observe that (\ref{diverge}) has the general form 
$\prod an^b$, then one write 
\bb 
\prod an^b = e^{\sum_n log s n^b} = e^{\sum_n log a + b\sum_n
log n}, \label{regu} 
\ee 
and using the Riemann $\zeta$-function 
\bb 
\zeta (s) = \sum_n {1\over n^s},  
\ee 
one see that 
\bn 
\prod an^b &=& e^{\log a \lim_{s\rightarrow 0} .
\sum_{n=1}^{n=\infty}  n^{-s} + b
\lim_{s\rightarrow 0} {d\over ds} \sum n^{-s}} \nonumber\\ 
&=& e^{\log a \zeta (0) + b \zeta^{'} (0)}. 
\en 

By analitic continuation one see that $\zeta (0) = -\12$ and
$\zeta^{'} (0) = -\12 log 2\pi$ and then 
\bb 
\prod an^b = a^{-\12} {(2\pi)}^{b\over 2}, \nonumber 
\ee 
so that (\ref{regu}) is simply $1/\Delta t$ and the propagator
becomes 
\bb 
\G = {1\over \sqrt{\Delta t}} e^{i{(\Delta x)}^2/\Delta t},
\nonumber 
\ee 
that is the standard result. 

In the previous problem we have assumed that the manifold is
defined on 
\bb 
-\infty < x < \infty. \nonumber 
\ee   

The next question is, what happens if the manifold has another
topological structure such as a circle or a torus etc.?. 

If the manifold has the topology of a circle, the boundary
conditions (\ref{bc}) do not define completely the problem and
one must modify (\ref{bc}) in the following way 
\bn 
x(t_1) &=& x_1 \nonumber \\
x(t_2) &=& x_2 + 2n\pi, \label{bc1} 
\en 
where $n$ is an integer number (winding number). 

A physical system described by the lagrangian 
\bb 
L = \12 {\dot x}^2, \label{lagra}
\ee 
with the boundary conditions (\ref{bc1}), is called quantum
rotator and it is the most simple example defined on a
multiply connected manifold. 

The strategy that we will follow below (and in essence is due to
Schulman) is to solve this example in detail and afterwards
to derive a general formula. 

Let us start making the identification in (\ref{lagra}) $x
\rightarrow \phi$ and (\ref{bc1}) becomes 
\bn 
\phi(t_1) &=& \phi_1 \nonumber \\ 
\phi(t_2) &=& \phi_2 + 2n\pi. \label{bc2} 
\en      

The propagation amplitude for this case becomes 
\bb 
 G_n[\phi_2, \phi_1]= \int \D\phi(t) e^{{i\over h} \int_1^2 dt 
\12 {\dot \phi}^2},
\label{formula1} 
\ee 
provided that the boundary condition (\ref{bc2}) are assumed. When 
(\ref{formula1}) is computed using (\ref{bc2}), the propagation
amplitude will depend on $n$, for this reason we
have written $G_n[\phi_2, \phi_1]$. 

One solve this problem in complete analogy with the free 
non-relativistic particle, in fact making the change of
variables 
\bn  
\phi(t) &=& \phi_1 + {{\Delta \phi + 2n\pi} \over \Delta t} (t -
t_1) + \psi (t)
\nonumber \\ 
&=& \phi_{cl} + \psi (t), \label{eq1}
\en   
with $\phi_{cl}$ the classical solution of the equation of
motion and, also by consistency, the quantum fluctuations
satisfying $\psi (t_1) = 0 = \psi (t_2)$. 

Replacing (\ref{eq1}) in (\ref{formula1})  
\bn 
G_n[\phi_2, \phi_1] &=& e^{{i(\Delta \phi +2n\pi)}^2\over \Delta t} 
\int \D \psi e^{i\int_{t_1}^{t_2} dt \12 \psi {\partial_t}^2 \psi} 
\nonumber \\ 
&=& e^{{i(\Delta \phi +2n\pi)}^2\over \Delta t} \det {(\partial_t^2)}^{-\12}. 
\label{schu} 
\en 

The determinant is computed as in the free non-relativistic
particle case and the result is $\Delta t$. Thus, (\ref{schu})
is 
\bb 
G_n[\phi_2, \phi_1] = {1\over \sqrt{\Delta t}} 
e^{{({\Delta \phi + 2n\pi})}^2\over
\Delta t} \label{final} 
\ee 

The expression (\ref{final}) is the propagation amplitude for a
fix homotopy class and, in consequence, the total propagation
amplitude is 
\bb 
G [\phi_2, \phi_1] = \sum_{n=-\infty}^{n=+\infty} \Xi_n G_n [\phi_2, \phi_1], 
\label{homo1} 
\ee 
where $\Xi_n$ is a factor that we must determine. By invoking
completness and unitarity of the Green function {\it i.e.} 
\bb 
\Xi^{\ast}_n \Xi_m = \Xi_{n+m}, \,\,\,\,\,\,\,\,\,\,\,\,\,\, 
\Xi^{\ast}_n \Xi_n = 1, \nonumber 
\ee 
one find that $\Xi_n$ must be $e^{n\delta}$ where $\delta$ is a
phase. 

Using the identity 
\bn 
\vartheta_3 (\tau, z) &=& \sum_{n=-\infty}^{n=+\infty} e^{2inz + i\pi
n^2 \tau} \nonumber \\ 
&=& {(-\tau)}^{-\12} e^{{z^2\over i\pi \tau}} \vartheta_3
({z\over \tau}, -{1\over \tau}), 
\en 
then, one find the final expression  
\bb 
G [\phi_2, \phi_1] = {1\over 2\pi} \exp [ i{{\delta \Delta \phi}\over 2\pi} 
- i{\delta^2 \Delta \over 8\pi^2} ] 
\vartheta_3 ( {\Delta \phi \over 2} - {{\Delta t \delta}\over
4\pi}, -{\Delta t\over 2\pi}), \label{sch} 
\ee 
that is the result found by Schulman in 1968, althought the
derivation given by him is slightly different. 

Finally we will discuss briefly the formal derivation of the
heuristic formula (\ref{homo}). 

Firstly one should note the existence of two equivalent
manifolds, ${\cal M}$ and their universal covering 
${\tilde {\cal M}}$. ${\cal M}$ is a multiply connected manifold while 
${\tilde {\cal M}}$ is simply connected. 

Both manifolds are related by 
\bb 
{\cal M} = {{\tilde {\cal M}} \over \Gamma}, \label{mani} 
\ee 
where $\Gamma$ is a discrete group. By definition the quocient 
${{\tilde {\cal M}} \over \Gamma}$ is the set of all 
homotopy classes, {\it i.e} 
\bb 
{{\tilde {\cal M}} \over \Gamma} = \{ [x], x\in {\tilde {\cal M}}\}. 
\ee  
Then, two point $x$ and $\tilde x$ are equivalent under $\Gamma$
if there exists an element $g\in \Gamma$ such that 
\bb 
{\tilde x}= x.g. \label{relation} 
\ee 

In the case considered above, ${\cal M} = S^1$ and ${\tilde M} =
\Re$ and the relation (\ref{relation}) is 
\bb 
{\tilde x} = x + 2\pi n, 
\ee 
while $\Gamma = {\cal Z}$ where ${\cal Z}$ is the integer group.
Once this nomenclature is introduced one can define the path
integral. 

Let ${\tilde \psi } ({\tilde x})$ be the wave function on
${\tilde {\cal M}} = \Re$, this wave function is continuous and
univalued then (\ref{relation}) becomes 
\bb 
{\tilde \psi } ({\tilde x}.g ) = a(g) . {\tilde \psi } ({\tilde
x}), \forall g \in {\cal Z} 
\ee 
if one impose the normalization of the wave function  
\bb 
\vert a(g) \vert = 1, 
\ee 
then $a(g)$ is a phase that satisfies the following property;
let the wave function with a well defined value, then there exists the
pre-image ${\tilde x} = p^{-1} (x)$ of $x$ with $\psi (x) =
{\tilde \psi} ({\tilde x_0})$; after a complete turn around in
$S^1$, $\psi (x)$ takes another value and the
pre-image will be different, say ${\tilde x}.g_1$ then 
\bb 
{\tilde \psi} ({\tilde x_0}) \rightarrow 
{\tilde \psi} ({\tilde x_0}.g_1) = a(g_1). {\tilde \psi}
({\tilde x_0}). \label{wave} 
\ee 

Giving another turn around one find 
\bn 
{\tilde \psi} ({\tilde x_0}.g_1) \rightarrow 
{\tilde \psi} ({\tilde x_0}.g_1 g_2)&=& 
a(g_2) {\tilde \psi} ({\tilde x_0}.g_1) \nonumber \\ 
&=& a(g_1). a(g_2) {\tilde \psi} ({\tilde x_0}), 
\en 
and, as a consequence 
\bb 
{\tilde \psi} ({\tilde x_0}.g_1 g_2) = a(g_1). a(g_2) {\tilde
\psi} ({\tilde x_0}), 
\ee 
thus 
\bb 
a (g_1.g_2) = a(g_1). a(g_2).  
\ee 

This last equation tell us, again, that $a(g)$ is a phase 
but also that there is a close relation between phase factors and
the group. In the case at hand, the phase factor is an unitary
irreducible representation of ${\cal Z}$. 

The propagator in ${\tilde {\cal M}}$ is defined as usual, 
\bb 
{\tilde \psi} ({\tilde x}_1, {\tilde t}_1) 
= \int_{\tilde {\cal M}} d{\tilde x} \,\,\,{\tilde G}
[{\tilde x}_2, {\tilde t}_2; 
{\tilde x}_1, {\tilde t}_1]\,\,\, 
{\tilde \psi} ({\tilde x}_2, {\tilde t}_2) , \label{fey} 
\ee 
where ${\tilde G}$ is the propagator for univalued functions on
${\cal M}$ with space constituted by infinity copies of ${\cal
M}$. Assuming continuity on the univalued functions, then
one can write (\ref{fey}) as 
\bb 
{\tilde \psi} ({\tilde x}_1, {\tilde t}_1) 
= \sum_{\Re} 
\int_{\Re} d{\tilde x}_1 \,\,\,{\tilde G}[{\tilde x}_2, {\tilde t}_2; 
{\tilde x}_1, {\tilde t}_1] 
\psi ({\tilde x}_1.g),
t^{'}, \label{fey0} 
\ee 
denoting the arbitrary point ${\tilde \prime x}^{'} \in 
{\tilde {\cal M}}$ by ${\tilde x}^{'}_0= 
\{ {\tilde  x}^{'} g \}$ where $x^{'}_0$ belong to a
copy on ${\cal M}$ for some fundamental domain ${\cal M}_0$.
Thus 
\bb 
{\tilde \psi} ({\tilde x}, {\tilde t}) 
= \sum_{g\in {\cal Z}} 
\int_{{\tilde {\cal M}_0}.g} d({\tilde x}_0.g) 
{\tilde G}[{\tilde x}, {\tilde t}; 
{\tilde x}_0.g, {\tilde t}_1] {\tilde \psi} ({\tilde x}_0.g,
t^{'}), \label{fey1} 
\ee 
but as the copies are identical, the integration on any copy is the
same on the fundamental domain, then we write 
\bb 
{\tilde \psi} ({\tilde x}, {\tilde t}) 
= \sum_{g\in {\cal Z}} 
\biggl\{ \int_{{\tilde {\cal M}_0}} d({\tilde x}) 
{\tilde G}[{\tilde x}, {\tilde t}; 
{\tilde x}_0.g, {\tilde t}_1] \biggr\} {\tilde \psi} ({\tilde x}_0.g,
t^{'}), \label{fey2} 
\ee 
or 
\bb 
{\tilde \psi} ({\tilde x}, {\tilde t}) 
=  \int_{{\tilde {\cal M}_0}} d({\tilde x}) \sum_{g\in {\cal Z}} 
{\tilde G}[{\tilde x}_2, {\tilde t}_2; 
{\tilde x}_0.g, {\tilde t}_1] {\tilde \psi} ({\tilde x}_0.g,
t^{'}), \label{fey3} 
\ee 
  
Following these arguments, then if $\psi (x,t)$ is the wave
function in the point $x$, then exist a pre-image ${\tilde x}$
of $x$ where ${\tilde \psi} ( {\tilde x}, t) = \psi (x, t)$.
Furthermore, ${\cal M}$ and ${\tilde {\cal M}}$ are locally
homeomorphic  and, in consequence, $d{\tilde x} = dx$. Thus, 
\bb 
\psi ( {\tilde x}, {\tilde t}) = \int_{{\cal M}} 
d x G[{\tilde x}, {\tilde t}; x, t] \psi (x,t), \label{fey4} 
\ee 
where 
\bb 
G[{\tilde x}, {\tilde t}; x, t] = \sum_{g \in {\cal Z}} 
{\tilde G}[{\tilde x}_0, {\tilde t}_0; 
{\tilde x}, t^{'}] a(g), 
\ee 
with $x=p({\tilde x}_0$ and $ x = p( x_0)$. Making 
$x_0 \rightarrow x_0 g^{-1}$ and ${\tilde x}_0 \rightarrow 
{\tilde x}_0 g^{-1}$ and afterwards $g \rightarrow g^{-1}$,
we arrive finally  
\bb 
G[{\tilde x}, {\tilde t}; x, t] = \sum_{g \in {\cal Z}}
a(g^{-1}) {\tilde G}[{\tilde x}_0, {\tilde t}_0; 
{\tilde x}, t^{'}], \label{last}
\ee 
that is the standard formula for the propagator in a multiply
connected manifold \cite{dewitt},\cite{dowker}. 
Althought (\ref{last}) was derived for a
particular topology is a formula valid for general cases. 

\section{Applications} 

In this section we will apply the formulas derived in the above
section to several problems such the AB effect including spin
(and their relativistic extensions) and anyons. 

\bigskip 

$\underline{\it The\,\,\, AB\,\,\, Effect}$ In the AB effect the
formal propagation amplitude is 
\bb 
\G = \sum_n \Xi_n \int_{(n)} \D x e^{iS_{free}}. \label{ab} 
\ee 

In order to compute (\ref{ab}) it is convenient to discretize as
follow 
\bn 
\G &=& \lim_{n \rightarrow \infty} {\biggl({\rho\over i\pi
\Delta t}\biggr)}^n \int_{-\infty}^{+\infty}
...\int_{-\infty}^{+\infty} \prod_{k=1}^{n-1} dx_k dy_k
\nonumber \\ 
&\times & \exp \biggl[ i\rho \sum_{j=1}^n \biggl( {{(x_j -
x_{j-1})}\over \Delta t}^2 + 
{{(y_j - y_{j-1})}\over \Delta t}^2 \biggr)\biggr], \label{ab1}
\en 
where $\rho = m/2, \Delta t = {(t_2 - t_1)\over m}$. 

Using polar coordinates, 
\bn 
x_j &=& r_j \cos \theta_j, \nonumber \\ 
y_j &=& r_j \sin \theta_j, 
\en 
and writing $dx_k dy_k = r_k dr_k d\theta_k$ and 
\bb 
{(x_j - x_{j-1})}^2 + {(y_j - y_{j-1})}^2 = r_j^2 + r_{j-1}^2 - 
r_j r_{j-1} \cos {(\theta_j  - \theta_{j-1})}, 
\ee 
equation (\ref{ab1}) becomes 
\bn 
\G &=& \lim_{n \rightarrow \infty} {\biggl({\rho\over i\pi
\Delta t}\biggr)}^n \int_{0}^{\infty}
...\int_{-\pi}^{+\pi} \prod_{k=1}^{n-1} r_k dr_k d\theta_k 
\nonumber \\ 
&\times & \exp \biggl[ {i\rho\over \Delta t} \sum_{j=1}^n
\biggl( r_j^2 + r_{j-1}^2 - 
r_j r_{j-1} \cos {(\theta_j  - \theta_{j-1})} 
\biggr)\biggr]. \label{ab2}
\en 

Now we impose that the electron can turn around the solenoid, technically 
this is equivalent to impose the constraint 
\bb 
\phi + 2\pi m - \sum_{j=1}^n (\theta_j - \theta_{j-1}) \nonumber 
\ee 
via a delta function, {\it i.e} 
\bn 
\G &=& \lim_{n \rightarrow \infty} {\biggl({\rho\over i\pi
\Delta t}\biggr)}^n \int_{0}^{\infty}
...\int_{-\pi}^{+\pi} \prod_{k=1}^{n-1} r_k dr_k d\theta_k 
\delta [\phi + 2\pi m - \sum_{j=1}^n (\theta_j - \theta_{j-1})]
\nonumber \\ 
&\times & \exp \biggl[ {i\rho\over \Delta t} \sum_{j=1}^n
\biggl( r_j^2 + r_{j-1}^2 - 
r_j r_{j-1} \cos {(\theta_j  - \theta_{j-1})} 
\biggr)\biggr], \label{ab3}
\en 
where $\phi$ is the angle between the source of electron, the center of 
the solenoid and the screen (see fig. 3). 
\begin{figure}
\vspace{5cm}
\caption{ Here $R$ and $R^{'}$ are the distances between the
source and the screen to the centre of the solenoid.} 
\end{figure}

Now one can exponenciate the $\delta$ function and after a tedious
calculation we find 
\bn 
&G&[x_2, x_1]_m = \lim_{n \rightarrow \infty} {\biggl({\rho\over i\pi
\Delta t}\biggr)}^n \int_{0}^{\infty}
...\int_{-\pi}^{+\pi} \prod_{k=1}^{n-1} r_k dr_k {1\over 2\pi} 
\int_{-\infty}^{\infty} d\lambda e^{i\lambda (\phi + 2\pi m)}   
\nonumber 
\\ 
&\times & \prod_{j=1}^n \exp \biggl[ {i\rho\over \Delta t} 
(r_j^2 + r_{j-1}^2 )\biggr] \int_{-\pi}^{\pi} ...
\int_{-\pi}^{\pi} \prod_{l=1}^{n-1} d\chi_l 
e^{-{2i\rho\over \Delta t}( r_j r_{j-1} \cos \chi_j - i\lambda
\chi_j)}, \label{ab4}
\en  

In order to compute these integrals we use the asymptotic formula
\bb 
\int_{-pi}^{+\pi} d\chi e^{i\lambda \chi + z\cos \chi} 
\rightarrow 2\pi I_{\vert \lambda\vert (z)} \label{asy} 
\ee 
valid in the limit $z\rightarrow \infty$. 

Integrating in $\chi$, and $r$  
\bb 
\G_m = {\rho\over i \pi \Delta t} e^{i{\rho\over \Delta t} (R^2
+ R^{{'}2})} \int_{-\infty}^\infty d\lambda e^{i\lambda (\phi +
2\pi m)} I_{\vert \lambda \vert} ( -2iRR^{'}/\Delta t),
\label{ab5} 
\ee 
thas is the the propagator for the $m$-th class of homotopy for
the AB effect. This formula was obtained by first time by
Inomata \cite{ino} and Gerry and Singh \cite{chris} in 1979 and 
simplified by Shiek\cite{shiek} more recently.

The total propagator is 
\bb 
\G = \sum_{n=-\infty}^{n=\infty} e^{2\pi i\alpha m} \G_m, \label{ab6}    
 \ee 
where $\alpha$ is the magnetic flux. Replacing (\ref{ab5}) in 
(\ref{ab6}) 
\bb 
\G = {\rho\over i\pi \Delta t} e^{i{\rho\over \Delta t} (R^2
+ R^{{'}2})} \sum_{-\infty}^{\infty} {(-i)}^{\vert m + \alpha
\vert} e^{-i(m + \alpha)\phi} J_{\vert m+ \alpha \vert} 
\biggl( {2RR^{'}\rho \over \Delta t}\biggl). \label{ab7} 
\ee
 
Equation (\ref{ab7}) has several \lq \lq sub-applications", as
was mentioned in section 2, the motion of two anyons is an
example about it, in fact let us consider the motion of $2$
free particles in a plane, the lagrangian is 
\bb 
L = \12 {\dot {\vec x}_1}^2 + \12 {\dot {\vec x}_2}^2, \label{lag1}
\ee 
and defining relative coordinates and the center of mass as usual, one have 
\bb 
L = \12 {\dot {\vec x}}^2 + \12 {\dot {\vec X}_{CM}}^2 
+ \alpha {d\Theta ({\vec x})\over dt}, \label{lag2}
\ee 
where ${\vec x}= {\vec x}_2 - {\vec x}_1$. The coordinate 
${\vec X}_{CM}$ is
the position of the center of mass, ${d \Theta\over dt}$ is a
topological that have been added by hand and that, classically,
does not contributes to the equation of motion and $\Theta$ is the relative 
angle between the particles. 

The partition function for this system becomes (the motion of
the center of mass is trivially decoupled) 

\bb 
Z = \12 \int d^2 {\vec r} \biggl[ <{\vec r}\vert e^{-\beta
H_{rel}} \vert {\vec r}> + 
<{\vec r}\vert e^{-\beta H_{rel}} \vert -{\vec r}> \biggr],
\label{two} 
\ee 
with  
\bb 
L_{rel} = \12 M{\dot {\vec r}}^2  + \alpha {\dot \theta }, 
\ee  

The brackets that appears in (\ref{two}) are just the
definition of the Green function and it was computed previously.
However the propagator is divergent and one must regularize 
the expression 
\bb 
Z = \sum_{n=-\infty}^{n= \infty} \int_0^\infty dx e^{-x}
I_{\vert {n - \alpha}\vert} (x). \label{sr} 
\ee 

In order to regularize one replace $e^{-x}$ by 
$e^{\varepsilon x}$ and take the limit 
$\varepsilon \rightarrow 0$ at the end of the calculation. 
The interested reader in the explicit calculations can see ref.
\cite{arobas}, the final result is 
\bn 
F_\nu (a) &= &\int_0^\infty dx e^{-ax} I_\nu (x) = {1\over
\sqrt{a^2 -1}} {\biggl[ a + \sqrt{a + \sqrt{a^2 -1}}
\biggr]}^{-\nu} \nonumber 
\\ 
F_\nu (1 + \varepsilon &)& \rightarrow {1\over
\sqrt{2\varepsilon}} {\biggl[ 1 + \sqrt{2}\varepsilon
\biggr]}^{-\nu}. 
\en 

With these expressions in mind one can compute the second virial
coefficient 
\bb 
B(\alpha, T) = 2 \lambda^2_T Z, \label{virial} 
\ee 
with $\lambda_T = {(2\pi h^2/ Mkt)}^{\12}$. 

If one expand around the Fermi statistic $\alpha = 2j + 1 +
\delta$ then one find 
\bb 
B ( \alpha = 2j + 1 + \delta, T) = {1\over 4} \lambda_T^2 + 2
\lambda_T^2. 
\ee 

More details about the calculation can be found {\it e.g} in the
Lerda's book \cite{lerda}. 
\bigskip 

$\underline{Relativistic\,\,\, Aharanov-Bohm\,\,\, Effect}$
\bigskip

The relativistic extension of the AB effect is straightforward,
but firstly one must define the path integral for a relativistic
particle\cite{polsch}.

A relativistic particle is defined by the following lagrangian
\bb 
L = {1\over 2N} {\dot x}^2 - \12 m^2 N, \label{ein} 
\ee 
where $N$ is the einbein. 

The classical symmetries of (\ref{ein}) are 
\bb 
\delta x^\mu = \epsilon {\dot x}^\mu,
\,\,\,\,\,\,\,\,\,\,\,\,\,\, \delta N = {\dot {(\epsilon N)}},
\label{si} 
\ee 

The next step consists in to compute the propagation amplitude
associated to (\ref{ein}), however it is not trivial because the
relativistic particle is a generally covariant system and the
propagator must be written \'a la Faddeev-Popov, {\it i.e.} 
\bb 
\G = \int \D N \D x^\mu \det (N)^{-1} \delta (f(N)) \det
({\delta f(N)\over \delta \epsilon}) \,\,e^{iS}. \label{rp}
\ee 

This expression deserves some explanations. Firstly we have
inserted the factor $\det (N)^{-1}$ by hand in order to have a
functional measure invariant under general coordinate transformations;
secondly the remaining factors are the usual terms of the
Faddeev-Popov procedure, being $f(N) =0$ the gauge condition. 

An appropiate gauge condition for this problem is ${\dot N} = 0$
(proper-time gauge) and having in account the causality
principle (\ref{rp}) becomes 
\bb 
\G = \int_0^\infty dT \int \D x^\mu \,\,\, e^{i\int_1^2 d\tau 
({1\over 2N(0)} {\dot x}^2 - \12 m^2 N(0))}, \label{gf}
\ee 
with $T=N(0)\Delta \tau$ and the boundary condition 
\bb 
x^\mu (\tau_1) = x^\mu_1, \,\,\,\,\,\,\,\,\,\,\,\, 
x^\mu (\tau_2) = x^\mu_2, \label{bbc} 
\ee 
has been assumed. The formula (\ref{gf}) was found by Schwinger
in 1951.

In order to compute (\ref{gf}) one repeat the arguments
given in the non-relativistic case, making the change of
variables 
\bn 
x^\mu (\tau) &=& x^\mu_1 + {\Delta x^\mu \over \Delta \tau}
(\tau - \tau_1) + y^\mu (\tau) \nonumber \\ 
&=& x^\mu_{cl} + y^\mu (\tau) \label{qf} 
\en
where $x^\mu_{cl}$ is the classical solution of the equation of
motion and $y^\mu$ is a quantum fluctuation that satisfies 
\bb 
y^\mu (\tau_1) = 0 = y^\mu (\tau_2). \label{qq}
\ee  

Replacing (\ref{qq}) in (\ref{gf}) one find 
\bn
\G &=& \int_0^\infty dT T^{-{D/2}} \,\,e^{i{{(\Delta x)}^2\over
2T} -i{m^2\over 2}T}, \nonumber \\
&=& \int {d^Dp\over {(2\pi)}^D} {e^{ip.\Delta x}\over {p^2 +
m^2}}, \label{kg}
\en 
that is the expected result. 

The next step is to apply these results in order to study the
relativistic AB effect. The main idea is simple, one write the
propagation amplitude for a relativistic particle as was
discussed above and afterwards one separate the vector $x^\mu$
in components $(x^0, x^1, x^2)$. 

The main steps are the following: 

i) Firstly instead of (\ref{gf}) one write 
\bn 
\G = \int_0^\infty &dT& 
e^{-i\12 m^2 N(0)} \times \nonumber 
\\ 
&\times&\int \D x^0 
e^{-i \int_1^2 d \tau {1\over 2N(0)} {({\dot x^0})}^2} 
 \,\,\,\int \D x e^{i\int_1^2 d\tau 
({1\over 2N(0)} {\dot x}^2)}, \label{gf1}
\en  
then in (\ref{gf1}) one can consider formally the integral in
$x_0$ as an ordinary free non-relativistic particle with mass 
$N_0^{-1}$ moving in a
one dimensional space, the result of this integration is trivial
\bb 
{1\over \sqrt{T}} \,\,\,e^{-i{{\Delta x_0}^2\over 2T}}.
\label{fpro} 
\ee

ii) The integral in the spatial coordinates is more complicated
but one can map this problem to a non-relativistic problem 
with formal mass $N_0^{-1}$. 

The final result is 
\bn 
G [x_2, x_1] &=& \lim_{\epsilon \rightarrow 0} \sum_{n= -\infty}^{+\infty} 
{{(-i)}^{\vert {n + \alpha} \vert}\over \vert {n + \alpha}} 
\,\,e^{-i{n+\alpha}\phi} \int {d^3p\over {(2\pi)}^3}
e^{ip.\Delta x} \times \nonumber 
\\ 
&\times& \biggl( J_{\vert {n + \alpha +1}\vert} (\sqrt{RR^\prime}\rho) + 
J_{\vert {n + \alpha -1}\vert} (\sqrt{RR^\prime}\rho) \biggr) 
\times \nonumber 
\\  
&\times&\biggl( K_{\vert {n + \alpha +1}\vert} (\sqrt{RR^\prime}\rho) + 
K_{\vert {n + \alpha -1}\vert} (\sqrt{RR^\prime}\rho) \biggr), 
\label{exact} 
\en
with $\rho = \epsilon - (p^2 + m^2)$. 

For details and other applications see \cite{gr}.

\section{ Anyons in Two Dimensions}

In this section we will discuss the idea of anyon from a more
general point of view, but before let us consider a particular case known as
Bose-Fermi Transmutation (BFt). 

The idea due to Polyakov \cite{poli} consist in to take an
spinning particle described by the action\cite{dhar}
\bb 
S = \int d\tau ( m\sqrt{{\dot x}^2} - {i\over 2} \theta_\mu
{\dot \theta}^\mu -  {i\over 2} \theta_5
{\dot \theta}_5 +\lambda \theta^\mu {\dot x}_\mu 
+ \sqrt{{\dot x}^2} \lambda \theta_5 ), \label{dhar} 
\ee 
with $\theta_\mu , \theta_5$ and $\lambda$ fermionic variables.
Then when one integrate
the fermionic variables one find a bosonic description
of a spinning particle or more precisely, an action
like 
\bb 
S = \int d\tau ( m\sqrt{{\dot x}^2} + topological \,\,\,
invariant), \label{pd} 
\ee

We will more precisely this result below.

In order to define appropriately the path integral one start
defining the gauge condition 
\bb 
\theta_5 = 0, \label{gauge} 
\ee 
that is consistent with the constraint
$\theta^\mu {\dot x}_\mu = 0$. 

The next step consists proposing the decomposition for the
fermionic variable 
\bb 
\theta^\mu = n^\mu_1 \kappa_1 + n^\mu_2 \kappa_2 + e^\mu
\kappa_e, \label{deco} 
\ee 
where $n_1, n_2$ and $e$ are tri-vectors that satisfy 
\bb 
e^\mu = {{\dot x}_\mu \over \sqrt{{\dot x}^2}},
\,\,\,\,\,\,\,\,\,\, n^\mu_1.e_\mu = 0, \,\,\,\,\,\,\,\,\,\,\,\,
n^\mu_i.n_{\mu j} = -\delta_{ij}.  \label{n} 
\ee 

The decomposition is equivalent to choose a Frenet-Serret frame
where the $n$'s are the normal vector and $e$ is the vector
tangent to the worldline. 

The effective fermionic action is computed from 
\bn 
e^{iS(x)} &=& \int \D \theta^\mu \D \theta_5 \D \lambda \,\, 
\delta (\theta_5 ) \,\, e^{i S} \nonumber 
\\ 
&=& \exp \bigg[ im \int \tau \sqrt{{\dot x}^2}\biggr]  
\Phi \label{effective} 
\en 
where $\Phi$ is the Polyakov spin factor defined as 
\bn 
\Phi &=& \int \D \kappa_1 \D \kappa_2 \exp \biggl[ \int d\tau 
\biggl( -\12 (\kappa_1 {\dot \kappa}_1 + \kappa_2 {\dot
\kappa}_2) + (n_1.{\dot n}_2) \kappa_1 \kappa_2 \biggr)\biggr]
\nonumber 
\\ 
&=& \det \biggl[ {d\over d\tau} + (n_1.{\dot n}_2) \biggr].
\label{de} 
\en 

The calculation of the determinant is
straightforward\cite{suecos},\cite{cortes} 
\bb 
\det \biggl[ {d\over d\tau} + (n_1.{\dot n}_2) \biggr] = 
e^{ic \int d\tau (n_1.n_2)} \cos \biggl[ \12 \int d\tau
(n_1.n_2) \biggr], \label{calculo} 
\ee
where c parametrize the differents possible regularizations. If one impose 
invariance under the interchanges $n_1$ and $n_2$ one find that
$c=0$ and the spin factor becomes 
\bb 
\Phi = \exp \biggl[ {i\over 2} \int dt (n_1.{\dot n}_2) \biggr] + 
\exp \biggl[ -{i\over 2} \int dt (n_1.{\dot n}_2) \biggr], \label{psf}
\ee 
where the factors $\pm \12$ denote the two possible spin
states. 

The next question is, how to generalize this result for other
spin?. The answer can be obtained from Chern-Simons theories,
let us start considering a set of $N$ relativistic particles minimally
coupled to an abelian Chern-Simons field, the action is 
\bb 
S = \sum_{k=1}^N m \int d\tau \,\,\sqrt{{\dot x}_k^2} + 
\int d^3x J_\mu A^\mu + {1\over 2\sigma} \int d^3x 
\epsilon^{\mu \nu \rho} A_\mu \partial_\nu A_\rho, \label{cs} 
\ee 
where $J^\mu = \sum_{k=1}^N x^\mu \delta^{(3)} (x - x_k(\tau))$. 

Then, one integrate the $A_\mu$ field and the result gives the
effective action 
\bb 
S_{eff} = m \int d\tau \,\,\sqrt{{\dot x}^2} - {\sigma\over 2}
\int d^3x d^3y J^\mu (x) K_{\mu \nu} (x,y) J^\nu (y), \label{eff}
\ee 
where $K_{\mu \nu} (x,y)$ is the inverse of the operator 
$\epsilon^{\mu \nu \rho} A_\mu \partial_\nu A_\rho$ and, of
course, satisfy 
\bb 
\epsilon^{\mu \nu \rho} A_\mu \partial_\nu A_\rho K_{\rho
\sigma} (x,y) = \delta_\sigma^\mu \delta (x - y),
\label{inverse} 
\ee 
replacing $J^\mu$ in (\ref{eff}) one find that the non-local
term becomes 
\bb 
 -{\sigma\over 2} \sum_{i,j = 1}^N I_{ij}, \label{gauss1} 
\ee 
where $I_{ij}$ is 
\bb 
I_{ij} = {1\over 4\pi} \int dx^\mu_i dx^\nu_j
\epsilon_{\mu \nu \rho} {{(x_i - x_j)}^\rho \over {\vert x_i -
x_j \vert}^3}. \label{I} 
\ee  

For closed curves $i j$, $I_{ij}$ becomes the linking number
while for $i=j$ there are additional contributions in the one
particle sector. These diagonal terms are computed\cite{stefano}
by a regularization as limit of non-diagonal terms, however the
result can be dependent of the regularization. In order to
perform this calculation one consider two curves infinitesimally
near, $I$ become 
\bb 
I = {1\over 4\pi} \int d\tau d\tau^\prime \lim_{\epsilon
\rightarrow 0} \epsilon_{\mu \nu \rho} 
{dx^\mu_\varepsilon \over d\tau} 
{{(x_\varepsilon (\tau) - x(\tau^\prime) )}^\nu \over 
{\vert x_\varepsilon (\tau) - x(\tau^\prime) \vert}^3} 
{dx^\rho_\varepsilon \over d\tau^\prime}, \label{curve}
\ee
with 
\bb 
x_\varepsilon (\tau) = x(\tau) + \varepsilon n(\tau), \label{ss}
\ee 
the non-conmutativity between the limits $\varepsilon \rightarrow
0$ and the integration, imply 
\bn 
T &=& \lim_{\varepsilon \rightarrow 0} L_\varepsilon - I
\nonumber \\ 
&=& {1\over 2\pi} \int d\tau\,\, \epsilon_{\mu \nu \rho} e^\rho
n^\nu {\dot n}^\rho, 
\en
where $e^\mu = {e^\mu \over \vert e \vert}$ is the normal
principl vector. The quantity $T$ is called the torsion of the
curve and $I$ is the self-linking of the curve. 

The difference $T -L$ is denoted by ${\cal W}$ and is the
writhing number number or cotorsion. Thus, the effect of the
Chern-Simons field is to produce the interaction lagrangian 
\bb 
L_{int} = s{\cal W}, \label{la}
\ee 
where $s = {\sigma\over 4\pi}$ is the spin of the system. From
this way one see that the BFt procedure is a particular case of
a more general formulation coming from of a Chern-Simon contruction.

Other applications of anyons in two dimensions are discussed in
the Fradkin's Lectures in this volume.

\section{Anyons in One-Dimension} 

In this section we will discuss the possibility of anyons in one
dimension. This possibility can be analized in complete analogy
with the two dimensional case; in two dimensions there are
anyons because have been points of the manifold. In one
dimension one can repeat the same arguments as follow; let
consider two non-relativistics particles moving on a line. 
For this system the configuration space is multiply connected (the real
line minus the origin) because the point where the particles collide is
singular. Classically the particles cannot go through each other, bouncing
off elastically every time they meet. Thus, the action of this system is
defined on the half-line [3,5], i.e. 
 
\begin{equation} 
S~=~ \int^{t_2}_{t_1} dt\,{1\over 2} {\dot x}^2,
\,\,\,\,\,\,\,\,\,\,\,\,\,\,\,\,\,\,\, (0<x< \infty), 
\label{lagrang}
\end{equation} 
where $x$ is the relative position of the two particles. As it is well
known, the Hamiltonian associated to (\ref{lagrang}) is not self-adjoint
on the naive Hilbert space because there is no conservation of 
probability at $x=0$. The Hamiltonian for (\ref{lagrang}), however, can 
be made self-adjoint by adopting a class of boundary conditions for all 
the states in the Hilbert space of the form \cite{von}\cite{simon} 

\begin{equation} 
\psi^{'} (0) = \gamma \psi (0), 
\label{condition}
\end{equation} 
where $\gamma$ is an arbitrary real parameter\footnote{There is
an alternative approach to this problem. One could include in
the classical configuration space the exchanged states where
$x<0$ is also permisible. The resulting system is described by
the same action as (1) but with $x\neq 0$ instead of $x>0$. In
this configuration space the self-adjoint extension of the
Hamiltonian imposes a condition that replaces (2), with two
complex parameters $\gamma_\pm $ instead of one $(\gamma)$. Here
we shall not follows this approach. It is remarkable however
that even if in our approach particle exchange is not included 
{\it ab initio}, quantum mechanics brings it in at the end.} 

The calculation of the propagator to go from an initial relative position 
$x_1$ to a final one $x_2$ for the above problem gives \cite{sharp}, 
\cite{fahri}

\begin{equation}
G_\gamma [x(t_2), x(t_1)] = G_0 (x_2 - x_1) + G_0 (x_2 + x_1) 
- 2\gamma \int_0^\infty d\lambda e^{-\gamma \lambda} 
G_0 (x_2 + x_1 + \lambda), 
\label{full}
\end{equation}
where $G_0$ is the Green function for a free non-relativistic particle, 
{\it i.e} 

\begin{equation}
G_0 (x - y) = {1\over \sqrt{2\pi i t}} {\st e}^{i{(x - y)}^2/
2t}. \label{free}
\end{equation}

Although in one spatial dimension it is  not possible to rotate particles, 
they can be exchanged and their \lq \lq spin" and 
statistics can be determined by the (anti-) symmetry of the wave 
function. This (anti-) symmetry in turn depends on the values of the 
parameter $\gamma$.

This last fact can be seen by taking the limits $\gamma = 0$ 
and $\gamma = \infty$ of (3) \cite{fahri}

\begin{equation}
G_{\gamma= 0, \infty} = G_0 (x_2 - x_1) \pm G_0(x_2 + x_1), 
\label{b-f}
\end{equation}

Under exchange of the positions of two particles in initial or final states, 
$G_{\gamma=0}$ is even and $G_{\gamma =\infty}$ is odd. Thus, for 
$\gamma = 0$ ($\gamma = \infty$) and the particles behave as bosons 
(fermions). The cases $0 < \gamma <\infty$ give particles with fractional 
spin and statistics \cite{leinaas}.

The propagator (\ref{full}) can also be obtained in the path integral 
representation, summing over all paths $-\infty < x(t) < \infty$, but in the 
presence of a repulsive potential $\gamma \delta(x)$. This problem was 
considered in\cite{sharp} -\cite{fahri} and the result is
 
\begin{equation} 
G_\gamma [x(t_2), x(t_1)] ~
=~ \int {\cal D} x(t) \,\,e^{iS}, 
\label{prop1}
\end{equation} 
with 
\begin{equation} S~=~ \int_{t_1}^{t_2} dt ~\bigg(\12{\dot x}^2 + \gamma 
\delta (x(t)) \biggr), 
\label{act1} 
\end{equation} 
Here $\D x(t)$ is the usual functional measure. The potential term 
$\gamma~ \delta(x(t))$ can be interpreted as a semi-transparent 
barrier at $x=0$ that allows the possibility of tunneling to the 
other side of the barrier. This is just another way of expressing 
the possibility of interchanging the (identical) particles. 
 
It is also interesting to note here that although in (1+1) dimensions 
the rotation group is discrete and the definition of the spin is a 
matter of convention, one may nevertheless view the one-dimensional 
motion on the half-line as a radial motion with orbital angular 
momentum $l=0$ \cite{grosche} in a central potential. This gives 
rise to another possible definition of spin by taking the following 
representation for the $\delta$-function
 
\begin{equation} \delta (x) = \lim_{\epsilon \rightarrow 0} 
{\sqrt{\epsilon}\over {x^2 + \epsilon}}.  
\end{equation}
 
Making a series expansion around $\epsilon =0$, the leading term 
$\gamma\sqrt{\epsilon}/x^2$ is analogous to the centrifugal potential for the 
radial equation in a spherically symmetric system, with 
$\sqrt{\epsilon}\gamma$ playing the role of an intrinsic angular momentum 
squared. Thus the spin of the system ($s$) can be defined by 
 
\begin{equation} 
s^2 = \sqrt{\epsilon}\gamma. 
\label{5}
\end{equation}
 
For real $s$ (\ref{5}) only makes sense when $\gamma>0$. This definition 
is consistent with the bosonic limit $\gamma =0$. For the fermionic case, 
the limit $\gamma = \infty$ mentioned above is to be interpreted as 
simultaneous with the limit $\epsilon  \rightarrow 0$ so that 
$\sqrt{\epsilon} \gamma =1/4$. It is in this sense that the non-relativistic 
quantum mechanics on the half-line describes one-dimensional anyons. However, 
the normalization $s=1/2$ for fermions is conventional. One can
extend these results for relativistics anyons although the
calculations are more involved only we will give the final
result for the propagator 
\begin{eqnarray}
G_\gamma [X(\tau_b), X(\tau_a)] &=& G_0 [X(\tau_b) - X(\tau_a)] + 
G_0 [X(\tau_b) + X(\tau_a)] \nonumber 
\\ 
&-& 2\gamma \int_0^\infty d\lambda e^{-\gamma \lambda} 
G_0 [X(\tau_b) + X(\tau_a)+ \lambda].
\label{full'}
\end{eqnarray}

The details are discussed in \cite{jj}. 

\section{Conclusions} 

In these lectures we have discussed several aspects of quantum
mechanics defined on non-trivial manifolds and, in particular,
anyons in one and two dimensions. During the conference we
discussed also other applications in one dimension and the
connection between these results and bosonization\cite{jvj}.

I would like to thanks the organizers of the conference, M. A. P\'erez,
A. Fernandez, C. Ram{\'{\i}}rez and J. C. D'Olivo for the kind
invitation for to visit Puebla, M\'exico. I thank also V. O.
Rivelles by several discussions. This work was partially
supported by grants FONDECYT 1950278 and DICYT.

\end{document}